\documentclass[%
 aps,
 amsmath,amssymb,
 reprint,%
]{revtex4-1}

\usepackage{amsmath}
\usepackage{amssymb}

\usepackage[dvips]{graphicx}
\usepackage{color}
\usepackage{textcomp}
\usepackage[T1]{fontenc}
\usepackage{dcolumn}
\usepackage{bm}
\usepackage{longtable}
\usepackage{bbm}

\begin{document}


\preprint{jcp\_v01}

\title{Unfolding Hidden Barriers by Active Enhanced Sampling}

\author{Jing Zhang}
\email{jing.zhang\texttt{@}kla-tencor.com}
\affiliation{KLA-Tencor, AI Division}

\author{Ming Chen}
\email{mingchen.chem\texttt{@}berkeley.edu}
\affiliation{Department of Chemistry, University of California, Berkeley}

\date{\today}

\begin{abstract}
Collective variable (CV) or order parameter based enhanced sampling algorithms have achieved great
success
due to their ability to efficiently explore the rough potential energy landscapes of complex systems.
However, the degeneracy of microscopic configurations, originating from the orthogonal space
perpendicular to the
CVs, is likely to shadow ``hidden barriers'' and greatly reduce the efficiency of
CV-based sampling. Here we demonstrate that systematic machine learning CV, through enhanced
sampling, can iteratively lift such degeneracies on the fly. We introduce an active learning scheme that consists
of a parametric CV learner based on deep neural network and a CV-based enhanced sampler. Our active
enhanced sampling (AES) algorithm is capable of identifying the least informative regions based on a historical
sample, forming a positive feedback loop between the CV learner and sampler. This approach is able to
globally preserve kinetic characteristics by incrementally enhancing both sample completeness and CV
quality.
\end{abstract}
\pacs{Valid PACS appear here}
\maketitle

Molecular dynamics (MD) simulations are an essential tool to understand the equilibria 
and kinetics of complex systems and processes, such as protein folding\cite{Rizzuti2013128}, 
drug binding~\cite{doi:10.1021/acs.jmedchem.5b01684}, 
phase transitions~\cite{doi:10.1021/acs.chemrev.5b00744}, glass states
~\cite{massobrio2015molecular,RevModPhys.83.587}, etc. Sampling equilibrium 
states and conformational changes requires the exploration of a ``rough'' high-dimensional
potential energy surface (PES), on which 
stable configurations are separated by relatively high barriers. This leads 
 to an exponential growth of equilibration time in a MD simulation. 
To avoid trapping in local minima, various enhanced sampling methods have been 
proposed to improve the sampling efficiency~\cite{Sugita1999141,Laio01102002,PhysRevLett.100.020603,
Maragliano2006168,doi:10.1021/jp805039u,doi:10.1063/1.2829861}.
One family of these methods including 
umbrella sampling~\cite{TORRIE1977187}, 
metadynamics~\cite{Laio01102002}, temperature accelerated MD~\cite{Maragliano2006168}, etc., 
forces the exploration of low-probability states via a biasing of the probability 
distribution of select degrees of freedom (DOF). 
Such DOFs are referred to as collective variables (CVs), which coarse-grain the high dimensional 
PES to a low dimensional free energy surface (FES). 

An ideal set of CVs should retain the kinetic characteristics 
of the system~\cite{doi:10.1063/1.475393,doi:10.1021/jp984837g,E2005242} on the FES, 
which requires that the CVs precisely describe the low 
free energy regions, especially critical transition paths between minima~\cite{E2005242}. 
Determining a small number of CVs to globally preserve 
kinetic information is quite challenging, 
due to the non-uniform intrinsic dimensionalities~\cite{doi:10.1063/1.3569857} 
and non-linear local 
structures of these regions~\cite{PhysRevLett.98.028102}. 
One natural approach for CV selection, which has achieved
some successes~\cite{Nymeyer18012000,doi:10.1021/ct900202f,
PhysRevLett.107.015701}, seeks to empirically construct CVs based
on physical intuition and structure characteristics. 
Other efforts have been focused 
on determining or training CVs through dimension reduction on simulation 
data~\cite{JCC:JCC3,Das27062006,Ceriotti09082011,doi:10.1137/070696325,
Tiwary15032016,doi:10.1063/1.4811489}. 
The resulting CVs from both approaches are often kept \textit{static}
throughout the entire enhanced sampling process. 

For complex chemical systems, the static form of the CVs usually leads to problematic degeneracies. 
In the space 
orthogonal to the CVs~\cite{Zheng23122008}, 
potential energy barriers, 
a.k.a. ``hidden barriers'', 
can separate important stable configurations. 
The transitions over hidden barriers that are shadowed by the chosen CVs 
are not observable on 
the FES. This phenomenon is called ``orthogonal space degeneracy''. 
When exploring the CV space, enhanced sampling algorithms only enhance 
the sampling of barrier crossing on the FES, while leaving transitions over hidden 
barriers unaffected. 
Therefore, enhanced sampling algorithms rely on CV selection methods to 
provide a set of less-degenerate CVs. Theoretically, the set of less-degenerate 
CVs can be constructed given either a prior understanding of the system~\cite{doi:10.1021/ct900202f,
PhysRevB.28.784,doi:10.1021/ct4002027} 
or a complete sampling of the system~\cite{Das27062006,
Ceriotti09082011,doi:10.1137/070696325,Tiwary15032016,doi:10.1063/1.4811489}. 
Yet in practice, 
it is very difficult to obtain this information in 
a finite amount of simulation time. Hence, to break degeneracy in orthogonal space, it is vital to 
establish a systematic and on-the-fly approach to CV construction for 
enhanced sampling algorithms.  

\begin{figure}
  \includegraphics{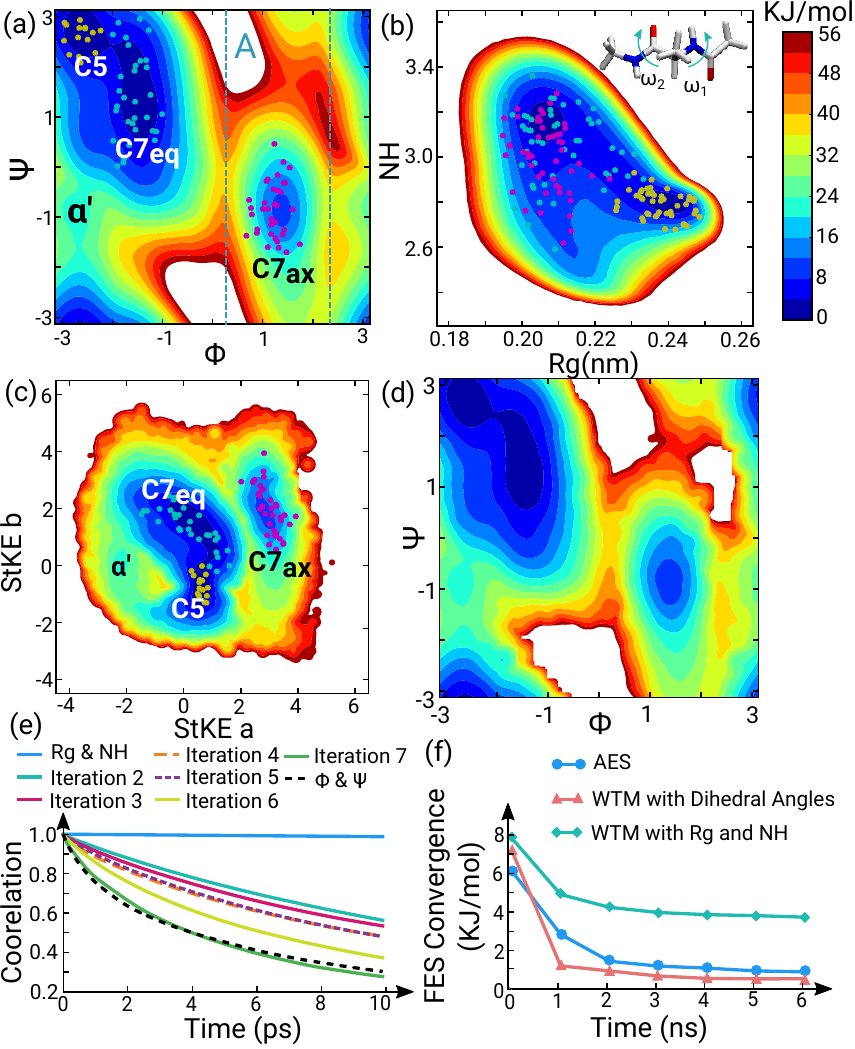}
  \caption{FESs of alanine dipeptide (b) in gas phase with respect to
(a) Ramachandran dihedral angles
($\Phi$ and $\Psi$), (b) Rg and NH are shown. Each FES was constructed from 
configurations sampled by 100ns WTMs with individual set of CVs. 
Four minima can be located on the (a) FES of $\Phi$ and $\Psi$, 
named as C7$_{eq}$ ,$C5$, C7$_{ax}$, and $\alpha'$.
AES was performed on this system
with $\mathbf{q}^h$ combining both Ramachandran dihedral angles and dihedral
angles of amide groups ($\omega_1$ and $\omega_2$) (inset of (b)). 
After 7 iterations (6.5ns in total) 
of AES, all configurations are accumulated and unbiased. These unbiased 
configurations are mapped onto the StKE CVs trained in AES iteration 7 to 
generate the FES shown in part (c). 
Configurations from C7$_{eq}$, C5, and C7$_{ax}$ are scattered on all three FESs as
cyan, yellow, magenta dots, respectively. 
Different $C(t)$ are calculated from WTM trajectories using Rg and NH,
StKE CVs trained in each AES iteration,
and Ramachandran dihedral angles. Each WTM was 100ns and
samples for calculating $C(t)$ were collected after 50ns such that biasing
potentials were well-converged. (f) FES convergence was calculated by 
unbiased samples from AES, WTM with $\Phi$ and $\Psi$, and WTM with 
Rg and NH. The benchmark FES is from part (a). 
\label{fig:aladi-conv}}
\end{figure}

Before explaining the methodology of AES, it is worthwhile to illustrate orthogonal 
space degeneracy by example. For this purpose the alanine dipeptide molecule was selected. 
As shown in Fig.~\ref{fig:aladi-conv}, two Ramachandran 
dihedral angles $(\Phi,\Psi)$ are usually considered as proper CVs to map all 
stable configurations~\cite{doi:10.1063/1.3569857} of the alanine dipeptide. 
Part (a) shows three major basins ($C7_{eq}$, $C5$ and $C7_{ax}$) on the 
FES of $\Phi$ and $\Psi$.
As a comparison, only two minima were located when the radius of gyration (Rg) 
and number of hydrogen bond 
(NH) (commonly employed when mapping biomolecule conformations
~\cite{doi:10.1021/jp805039u,Barducci03122013,Yu18102016}, see \textit{SI})
were selected as CVs, see Fig.~\ref{fig:aladi-conv}(b). 
It is clear that $C7_{eq}$ and $C7_{ax}$ are                
degenerate with similar Rg and NH values, which greatly reduces
the sampling efficiency of $C7_{ax}$, as illustrated in Fig.~\ref{fig:aladi-conv}(e). 
Denoting the $C7_{ax}$ basin as $A$, we estimate the sampling efficiency 
via a normalized 
auto-correlation function $C(t)=\langle\mathbbm{1}_A(\mathbf{x}(0))
\mathbbm{1}_A(\mathbf{x}(t))\rangle/\langle\mathbbm{1}_A^2(\mathbf{x}(0))\rangle$, 
where $A = \{\mathbf{x}:17.2^\circ\leq\Phi(\mathbf{x})\leq126.1^\circ\}$ 
and $\mathbf{x}$ are
system Cartesian coordinates.
$C(t)$ measures the possibility of finding the system, with initial C7$_{ax}$ configuration, 
staying in $A$ at time $t$. Faster decay of $C(t)$ 
suggests a shorter average time for the system to escape from $C7_{ax}$, and vice versa. 
Fig.~\ref{fig:aladi-conv}(e) clearly shows that well-tempered metadynamics (WTM)
~\cite{PhysRevLett.100.020603} with Rg and NH produces a very 
slow decay of $C(t)$, indicating that transitions between $C7_{ax}$ and 
$C7_{eq}$/$C5$ are not enhanced due to degeneracy
while applying $\Phi$ and $\Psi$ removes the degeneracy. Thus 
$C(t)$ decays much faster. Unlike alanine dipeptide, 
in a more general scenario, constructing a small working set of such CVs is almost 
impossible without a thorough knowledge of the system. 

Instead of well chosen CVs from prior knowledge, this work proposes Active Enhanced Sampling 
(AES) as a solution, which can start with arbitrary CVs and iteratively improve CV quality via 
active learning. Active learning is a semi-supervised learning algorithm to conductively query 
samples or desired outputs from the current least informative regions (CLIRs) as new learning 
samples. Similarly, AES introduces Stochastic Kinetic Embedding (StKE) to generate the low 
dimensional CV representation that preserves kinetic information and determines the CLIRs 
including degenerate states. Such CV representation (created by StKE) along with a FES sampler 
guides the MD simulation to explore these CLIRs more efficiently; on the other hand, the 
configurations generated by the sampler improves the learning of StKE. 


The formalism of StKE is described as below. 
Assuming Cartesian coordinates of the system are $\mathbf{x}$, $N_h$ generalized
coordinates are denoted as $\mathbf{q}^h(\mathbf{x})$ and selected to characterize all slow modes.
CVs are defined as
$N_l<N_h$ functions of $\mathbf{q}^h$, i.e. $\{\mathbf{q}^l(\mathbf{q}^h(\mathbf{x}))\}$. 
Given a set of samples of generalized coordinates $\mathbf{s}^h_n=\mathbf{q}^h(\mathbf{x}_n) 
\in \mathbb{R}^{N_h}$,  
$n=1,2,...,N_s$ with $N_s$ as the total number of samples, 
and the associated Boltzmann probability $p(\mathbf{s}^h_n)$, 
StKE determines 
a projection of 
$f:\mathbf{s}_n^h \in \mathbb{R}^{N_h}\rightarrow\mathbf{s}_n^l \in \mathbb{R}^{N_l}$, such 
that the diffusion 
distance~\cite{doi:10.1137/070696325} 
between each pair of datapoints is optimally retained. 
StKE assumes the samples are generated from an implicit diffusion process. 
A Markov chain is defined on $\{\mathbf{s}^h_n\}$ with an unnormalized transition matrix as 
$L(\mathbf{s}^h_i, \mathbf{s}^h_j)=K(\mathbf{s}^h_i, \mathbf{s}^h_j)/\sqrt{p(\mathbf{s}^h_i)p(\mathbf{s}^h_j)}$
where $K(\mathbf{s}^h_i, \mathbf{s}^h_j)$ is a Gaussian kernel describing the 
Brownian motion transition probability
from datapoint $\mathbf{s}^h_i$ to $\mathbf{s}^h_j$ within a finite time step and 
$p(\mathbf{s}^h_i)$ is estimated via kernel density estimator. 
For unbiased samples $\{\mathbf{s}^h\}$ with weights $\{\omega\}$ from enhanced sampler,
normalizing $L$ generates proper transition matrix $M$, i.e. 
$M_{i,j} \equiv M(\mathbf{s}^h_i, \mathbf{s}^h_j)=\omega_j L(\mathbf{s}^h_i, \mathbf{s}^h_j)/D_i$ where 
$D_i = \sum_{j}\omega_j L(\mathbf{s}^h_i, \mathbf{s}^h_j)$ (for derivation, see SI). 
It has been proven that in the limit of an infinite number of samples, $M$ will weakly
converge to the generator of the diffusion process~\cite{COIFMAN20065} . 
Different from several well-established dimensionality reduction methods for CVs, including 
diffusion map~\cite{doi:10.1137/070696325,doi:10.1063/1.3569857},
SGOOP~\cite{Tiwary15032016}, tICA~\cite{doi:10.1063/1.4811489}, etc, which utilize 
low-rank approximation that truncates the number of CVs at a chosen spectral gap, 
StKE adopts the spirit of tSNE~\cite{tSNE} by applying Kullback-Leibler divergence 
to estimate the similarity of $M$ between the higher-dimension ($M^{high}$) and the
lower-dimension ($M^{low}$) for all pairs of datapoints,
\begin{equation}
C=\sum_i\left(\sum_j M_{i,j}^{high}\log\frac{M_{i,j}^{high}}{M_{i,j}^{low}}\right)
\;\;\ldotp
\label{eq:obj-stke}
\end{equation}
To ensure that StKE learns an explicit function form 
for the projection function $f$, 
we assume that $f$ can be approximated by a parametric model 
$F(\mathbf{s}^h;W)\approx f(\mathbf{s}^h)$, where $W$ is trainable parameters for model $F$. 
A parametrized normalized
transition matrix in lower dimension defined as 
$\tilde{M}^{low}_{i,j}(W)\equiv M^{low}(F(\mathbf{s}^h_i; W),F(\mathbf{s}^h_j; W))$ replaces 
$M_{i,j}^{low}$ in Eq.(~\ref{eq:obj-stke}), generating the objective function $C(W)$. 
Thus, $W$ can be learned by 
minimizing $C(W)$ and model $F$ can be trained via 
efficient stochastic gradient descent method. 

As the parametric model $F(\mathbf{s}^h;W)$ has to be 
differentiable so that the biasing force 
can be evaluated in an MD simulation with enhanced sampling methods, 
a neural network is a practical choice for this model, e.g. 
multilayer perceptron (MLP). 
MLP consists of
multiple fully-connected layers and non-linear activations to simulate the 
complex function form for the projection $f$. 
Since the neural network is differentible with respect to both input and output space, this 
enables an estimate of $\partial f(\mathbf{s}^h)/\partial \mathbf{s}^h$ by 
$\partial F(\mathbf{s}^h;W)/\partial \mathbf{s}^h$, which further allows samplers to estimate biasing forces on the fly. 

AES uses WTM as FES sampler, 
WTM fills the FES with a time-dependent biasing potential 
by depositing Gaussians on the fly along the simulation trajectory
~\cite{Laio01102002,PhysRevLett.100.020603}, 
where gaussian heights in WTM decrease as the FES fills up. 
In the long time limit, it has been proven that the biasing potential 
will eventually converge to 
the scaled inverse FES while the CV samples 
display a Boltzmann distribution at higher temperature $T+\Delta T$
~\cite{PhysRevLett.100.020603,PhysRevLett.112.240602}, where $T$ is 
the system temperature and $\Delta T$ is a parameter in WTM. 
Since decreasing Gaussian heights can 
generate a more equilibrium-like simulation trajectory, it is easier to unbias 
samples to the correct ensemble distribution in WTM~\cite{JCC:JCC21305,
doi:10.1021/jp504920s}. 

The protocol of AES is summarized as follows: 

(i) AES starts from a short WTM simulation with arbitrary CVs. The initial
set of $\{\mathbf{s}^h\}$ are collected and unbiased following the method in
~\cite{doi:10.1021/jp504920s} to generate sample weights $\{\omega\}$. 

(ii) These samples are then resampled by enforcing a minimal pairwise distance 
$r_c$ to create a sparse description in low free energy regions. 
Probabilities of the resampled points $\{\tilde{\mathbf{s}}^h\}$, i.e.
$\{p(\tilde{\mathbf{s}}^h)\}$ are calculated as a high temperature ($T_h\geq T$) Boltzmann
distribution in order to emphasize low probability regions. 
$\{\tilde{\mathbf{s}}^h\}$ 
and $\{p(\tilde{\mathbf{s}}^h)\}$ are then used to train StKE CVs $\mathbf{q}^l$.

(iii) $\{\mathbf{s}^h\}$ are then
mapped onto the updated StKE CVs to generate $\{\mathbf{s}^l\}$ with which an initial
biasing potential is generated, i.e.
$V_{init}(\mathbf{q}^l(\mathbf{x}))=\frac{kT\Delta T}{T+\Delta T}
\log(\sum_{i}w_ie^{-\|\mathbf{q}^l(\mathbf{x})-\mathbf{s}^l_i\|^2/
\sigma_{init}}+P_0)-E_0$ where $k$ is the Boltzmann constant. 
$P_0$ and $E_0$ are constant such that
$\min V_{init}=0$ and $\max V_{init}$
equals the maximum biasing potential from the last WTM simulation. This initial biasing
potential is then used in next WTM with $\mathbf{q}^l$ as CVs. 

Steps (i) to (iii) forms one AES iteration. 
$\{\tilde{\mathbf{s}}^h\}$ and $\{p(\tilde{\mathbf{s}}^h)\}$
are accumulated through all previous iterations to generate next CVs and the whole
history of$\{\mathbf{s}^h\}$ and $\{w\}$ is kept for updating $V_{init}$.
By explicitly 
forming the positive feedback loop between StKE and WTM, 
AES incrementally improves both sample completeness and 
CV quality through iterations. 


Two systems were used to demonstrate the effectiveness of AES: alanine dipeptide 
and met-enkephalin. 
The simulations were performed by GROMACS 5~\cite{Abraham201519} with
OPLS-AA~\cite{doi:10.1021/jp003919d}
force field and PLUMED 2~\cite{Tribello2014604} was used for WTM simulation.
In alanine dipeptide example, samples from a 100ns 
WTM simulation with $\Phi$ and $\Psi$ were used to construct a benchmark FES 
(Fig.~\ref{fig:aladi-conv}(a)) 
with four minima (C7$_{eq}$, C5, C7$_{ax}$ and $\alpha'$). 
Considering that two additional dihedral angles $\omega_1$ 
and $\omega_2$ are 
also important to capture configuration-change kinetics~\cite{Bolhuis23052000}, 
these two dihedral angles, together with $\Phi$ and $\Psi$, are used as $\mathbf{q}^h$ 
inputs to StKE (inset of Fig.~\ref{fig:aladi-conv}(b)) 
for generating two CVs (i.e., StKE a and b). 
AES started from a 500ps WTM with Rg and NH, and followed by AES iterations each with 1ns WTM. 
After 7 AES iterations, configurations were accumulated and unbiased, then mapped back to $\Phi$ 
and $\Psi$ to generate the FES (Fig.~\ref{fig:aladi-conv}(d)). This FES is highly consistent with 
the benchmark. All four minima 
are quantitatively sampled in AES with correct free energy 
values. Such consistency demonstrates that AES can generate Boltzmann distributed 
samples after unbiasing. The configurations from AES were also mapped 
onto the StKE CVs from the last iteration, generating a FES in 
Fig.~\ref{fig:aladi-conv}(c). 

As mentioned earlier, Rg and NH 
lead to degeneracy between C7$_{eq}$ and C7$_{ax}$. 
As shown in Fig.~\ref{fig:aladi-conv}(c), 
StKE CVs are able to completely separate 
samples from different clusters. StKE CVs from AES iterations were 
then used in WTM simulations to calculate $C(t)$. Faster decay of $C(t)$ 
with respect to the number of AES iterations 
indicates quality improvement of StKE CVs due to 
the increasing completeness of data samples. 
At AES iteration 7, $C(t)$ decays as fast as the benchmark, 
indicating that StKE CVs are as good as $\Phi$ and $\Psi$ for 
preserving kinetic information in alanine dipeptide. 
We estimated FES convergence by calculating $L_1$-distance per area between 
a test FES and the benchmark FES in $\Phi$ and $\Psi$ space. 
The test FES for AES is estimated by accumulating StKE samples and mapping 
them back to $\Phi$ and $\Psi$ space. As shown in Fig.~\ref{fig:aladi-conv}(f), 
AES achieves similar FES 
convergence as benchmark, indicating the ability of AES to boost 
sampling efficiency by iteratively improving CVs. On the other side, WTM 
with Rg and NH fails to achieve FES convergence comparable to benchmark. 

\begin{figure}
  \includegraphics{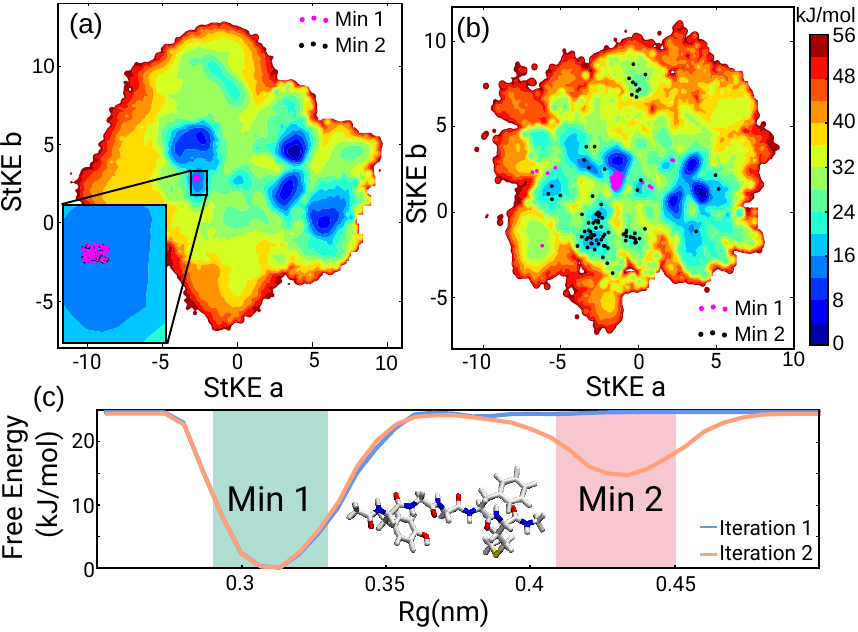}
  \caption{ 
Configurations sampled in AES 1st and 2nd iterations were mapped onto 
StKE CVs from iteration 1, generating FES shown in part (a). 
All the magenta and black dots lie in a
neighborhood around the bottom of one minimum highlighted in part (a). 
The magenta dots were chosen
so that their Rg values stay within the green block denoted as ``Min 1'' 
in part(c) while the black ones
were selected with their Rg values in the pink block denoted as ``Min 2''. 
Fixing the StKE CVs
at this minimum, the FES along Rg is also generated with unbiased samples from
iteration 1 (blue line) or with samples from both two iterations (pink line). 
It is clear 
that ``Min 2'' consists of new states sampled in iteration 2 with similar Rg values. 
In part (b), StKE CVs were trained with configurations
from both iteration 1 and 2, and samples corresponding to these configurations
were used to generate the FES in this part. Clearly,
the structures with respect to ``Min 2'' were separated out, forming different
minima on the FES. 
\label{fig:met-phys-cv}}
\end{figure}

Mapping both folded and unfolded conformations of peptide is another challenging 
problem~\cite{Chen17032015}. AES was tested on penta-peptide
met-enkephalin (Fig.~\ref{fig:met-phys-cv}(c)) in gas phase. 
StKE was used to embed 
10 Ramachandran dihedral angles to a 2D representation. AES was initiated 
by a 20ns WTM with Rg, NH and backbone-heavy-atom root-mean-square deviation 
(RMSD) as CVs, 
followed by 8 AES iterations each with 100ns WTM using 2D StKE CVs 
and Rg. 
The converged 3D FES and selected stable and metastable configurations 
are presented in SI.
In Fig.~\ref{fig:met-phys-cv}(a), we highlight the 
FES minimum where degeneracies occurred with the StKE CVs in 
iteration 1 due to incomplete sampling. When applying WTM with StKE CVs 
(from 1st iteration) and Rg, metastable configurations from degenerate states 
are discovered in iteration 2, as shown as the minimum 2 
in Fig.~\ref{fig:met-phys-cv}(c). 
After updating StKE CVs with these metastable configurations
from iteration 2, the degenerate stable states were separated as illustrated 
in Fig.~\ref{fig:met-phys-cv}(b), indicating that AES is capable of unfolding 
discovered hidden barriers from PES into FES that is defined by updated StKE CVs. 

\begin{figure}
  \centering
  \includegraphics{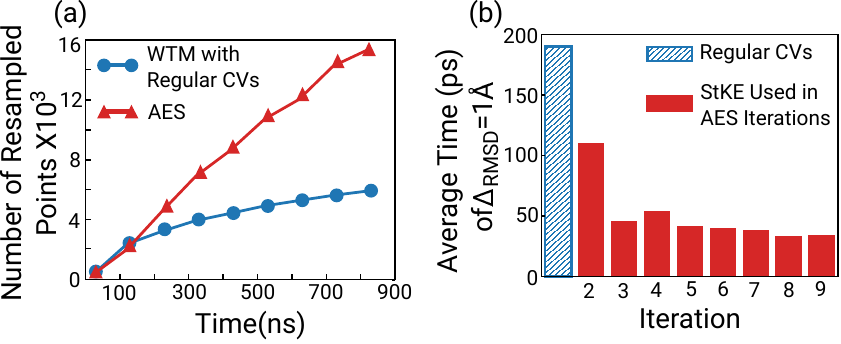}
  \caption{The number of resampled points in 10D Ramachandran dihedral angle space v.s.
total simulation time is recorded for both AES and metadynamics in part (a). This number
approximates the size of sampled conformational space due to the
lower bound on pairwise
distances between among these points. Part (b) records the average time in which
one structure changes to another with RMSD=1\AA. 
The average was taken from individual 400ns WTM with CVs used in each AES iteration. 
Only configurations sampled in the last 200ns were used to calculate the average, 
after biasing potential in WTM has filled low free energy regions in the first 200ns 
simulations. CVs used in iteration 1 are ``regular CVs''. 
\label{fig:met-conv}}
\end{figure}

To evaluate the efficiency of AES, a 1$\mu$s metadynamics simulation with 
Rg, NH and RMSD (``regular CVs'') was performed as comparison. 
As demonstrated in Fig.~\ref{fig:met-conv}(a), a faster 
exploration of the conformational space (represented as an increase in the resampled points)
is observed for AES. 
Beyond 200ns, AES is able to keep a high efficiency for discovering new configurations 
while the efficiency from WTM  
decreases. The linearly increasing of number of resampled points shows that 
unfolding discovered hidden barriers into FES by StKE encourages WTM to construct biasing 
potential more efficiently, while with {\it static}
CVs WTM spends majority of the simulation 
time revisiting the stable configurations. The ability of AES to guide the 
enhanced sampling simulations also significantly decreases the average time needed for 
conformational changes since majority of ``hidden barriers'' are removed in AES, as 
shown in Fig.~\ref{fig:met-conv}(b). 
In the long time limit, FES filling in WTM with regular CVs 
is unable to accelerate structural changes due to orthogonal space degeneracies, while  
StKE CVs remain faster conformational changes. 

Active enhanced sampling is a framework 
joining CV production and sampling to unfold discovered hidden barriers into FES. 
In analogy to active learning, on-the-fly training StKE CVs, together with the 
biasing potential in WTM, guides MD simulations to sample the CLIRs. Iteratively 
training StKE CVs promotes the removal of orthogonal space degeneracies, 
boosting the sampling efficiency 
comparing to WTM with static and/or human intuited CVs. 
In alanine dipeptide example, StKE retains 
both intra and inter cluster structures, more importantly, the kinetic information is
also preserved in StKE. 
In addition, 
minima on the FES are consistent with benchmark results, while 
human intuited CVs (i.e., Rg
and NH) are unable to identify all minima. 
In met-enkephalin system, AES demonstrates its ability to remove degeneracy on the fly,
leading to fast exploration of stable and metastable configurations and 
enhanced transitions.

Besides dihedral angles, other order parameters 
or those from dimension reduction algorithms, 
can be adopted as  $\mathbf{q}^h$ for StKE learning. 
Other than WTM, 
enhanced sampling methods, like temperature accelerated
molecular dynamics/driven 
adiabatic free energy dynamics~\cite{Maragliano2006168,doi:10.1021/jp805039u}, 
adaptive biasing force~\cite{doi:10.1063/1.2829861}, and unified free energy
dynamics~\cite{doi:10.1063/1.4733389} etc., can also be married with AES.
Although the present examples are calculated in gas phase, applying AES to condensed 
phase simulations is straightforward. 

We thank Mark E. Tuckerman, 
Yu Zhao and Tyler Y. Takeshita for reading the manuscript and giving suggestions. 
We also thank Phillip Geissler for useful discussion. 


%

\end{document}